\documentclass[showpacs,twocolumn,prb,aps,color]{revtex4}
\usepackage{graphicx}
\begin{document}

\title{Gapped spin liquid phase in the $J_{1}{-}J_{2}$ Heisenberg model by a
Bosonic resonating valence-bond ansatz}

\author{Tao Li,$^{1,2}$ Federico Becca,$^{2}$ Wenjun Hu,$^{2}$ and
Sandro Sorella$^{2}$}
\affiliation{
$^{1}$ Department of Physics, Renmin University of
China, Beijing 100872, P.R.China\\
$^{2}$ SISSA - International School for Advanced Studies and CNR-IOM
Istituto Officina dei Materiali, Democritos Center, Via Bonomea 265, 34136
Trieste, Italy
}

\date{\today}

\begin{abstract}
We study the ground-state phase diagram of the spin-$1/2$
$J_{1}{-}J_{2}$ Heisenberg model on the square lattice with an
accurate Bosonic resonating valence-bond (RVB) wave function. In
contrast to the RVB ansatz based on Schwinger Fermions, the
representation based on Schwinger Bosons, supplemented by a
variational Monte Carlo technique enforcing the exact projection
onto the physical subspace, is able to describe a fully gapped spin
liquid in the strongly frustrated regime. In particular, a fully
symmetric $Z_2$ spin liquid is stable between two antiferromagnetic
phases; a continuous transition at $J_{2}=0.4J_{1}$, when the
Marshall sign rule begins to be essentially violated, and a
first-order transition around $J_{2}=0.6J_{1}$ are present. Most
importantly, the triplet gap is found to have a non-monotonic
behavior, reaching a maximum around $J_{2}=0.51J_{1}$, when the
lowest spinon excitation moves from the $\Gamma$ to the $M$ point,
i.e., $\mathrm{k}=(\pi,0)$.
\end{abstract}

\maketitle

\section{Introduction}\label{sec:intro}

The search for quantum spin liquids in frustrated quantum antiferromagnets
has a long history.~\cite{RVB} In recent years, thanks to the advance of
numerical techniques, several candidates for spin liquids have emerged in
two-dimensional (2D) systems. These include the Hubbard model on the
honeycomb lattice,~\cite{honeycomb} the spin-$1/2$ Heisenberg model on Kagome
lattice,~\cite{kagome} and more recently the spin-$1/2$ $J_{1}{-}J_{2}$
Heisenberg model on the square lattice.~\cite{dmrg} In all these cases,
a small but finite spin gap has been found and, according to generalizations
of the Lieb-Schultz-Mattis theorem for higher dimensionalities,~\cite{hastings}
a topological degeneracy is expected. In spite of these results, descriptions
based upon Fermionic resonating valence-bond (RVB) theory predict more often
the existence of gapless spin-liquid states. For example, for the
$J_{1}{-}J_{2}$ model on the square lattice,~\cite{frvb} the Heisenberg model
on the triangular~\cite{yunoki} or the Kagome lattices,~\cite{kagomef} and,
more recently, also for the unfrustrated honeycomb lattice,~\cite{honey}
the Fermionic RVB theory always predicts a gapless spin liquid phase with
a Dirac-type spinon dispersion as the best variational state.

The $J_{1}{-}J_{2}$ model represents the simplest model to study the
effect of frustration in a (low-dimensional) magnetic system; for this
reason it has been investigated by many different approaches in the last
20 years.~\cite{schulz,sachdev,gelfand,sushkov,mambrini,darradi,verstraete}
At the classical level, the system is magnetically  ordered for
$J_{2}<0.5J_{1}$ with the standard antiferromagnetic pattern at
$\mathrm{q}=(\pi,\pi)$. For $J_{2}>0.5J_{1}$, the ordering wave vector is
moved to $\mathrm{q}=(\pi,0)$ or $(0,\pi)$; these two ordered phases are
separated by a first-order transition. Within the linear spin-wave approach,
which goes beyond the classical theory, quantum fluctuations destroy the
magnetic order in the intermediate region of
$0.4J_{1} \lesssim J_{2} \lesssim 0.6J_{1}$, hence leading to a magnetically
disordered state.~\cite{doucot}
However, the nature of this disordered phase is still elusive and several
proposals have been raised. These include valence-bond solids with broken
spatial symmetries~\cite{sachdev,mambrini,vbs1,vbs2} or gapless spin-liquid
states.~\cite{frvb} The latter proposal is especially attractive,
since it provides a simple and very accurate Fermionic RVB wave
function for $0.4J_{1} \lesssim J_{2} \lesssim 0.55J_{1}$. This
state has a Dirac-type spinon dispersion and $Z_2$ gauge structure
and becomes stable for $J_{2} \gtrsim 0.4J_{1}$.

More recently, density-matrix renormalization group (DMRG) calculations
provided some evidence for a fully gapped spin liquid in the intermediate
region of $0.4J_{1} \lesssim J_{2} \lesssim 0.62J_{1}$.~\cite{dmrg}
Within this numerical approach, the spin gap increases linearly from
$J_{2} \simeq 0.4J_{1}$, reaches a maximum around $J_{2} \simeq 0.59J_{1}$,
and then rapidly decreases. For $J_{2} \gtrsim 0.62J_{1}$, a collinear
magnetic order takes place. The spin-liquid phase determined by these DMRG
calculations is thus inconsistent with the Fermionic RVB theory, due to
presence of a finite spin gap.

In this paper, we investigate the spin-liquid phase of the $J_{1}{-}J_{2}$
model with a Bosonic RVB wave function.~\cite{liang} This is motivated by the
following reasons. First, while the Fermionic RVB state is found to be unable
to open a spin gap for this system, a Bosonic spin-liquid state is by
definition gapped, because otherwise the (Bosonic) spinon would condense
and the system would develop magnetic order. Second, since the spin-liquid
phase is found to exist in a quite small region between two magnetically
ordered phases (for which a Bosonic description is quite accurate), it is
natural to expect that the intermediate spin-liquid phase inherits some
Bosonic characteristic.

The Bosonic RVB state has been adopted in many previous
studies~\cite{sandvik,beach} and is found to describe quite well both the
magnetic ordered state and the disordered state for unfrustrated
systems.~\cite{xiu,bilayer} For frustrated magnetic systems, the use of the
Bosonic RVB wave function is very limited, since the loop gas algorithm for
the Bosonic RVB state encounters serious sign problems; moreover, the
computation of the wave function amplitude in the orthogonal Ising basis
involves permanents of matrices,~\cite{permanentbook} implying a computational
cost that grows exponentially with the size of the system. Only very recently,
this approach has been implemented on small clusters for the Kagome
lattice.~\cite{permanent}

Here, the Bosonic RVB state is obtained after projecting the ground state of
the mean-field Schwinger Boson Hamiltonian~\cite{schwinger} into the physical
subspace with one spin per site. After this projection, the wave function
turns out to be equivalent to the standard Liang-Doucot-Anderson RVB
ansatz,~\cite{liang} defined only in terms of a bosonic pairing function
(that connects opposite sub-lattices). To enforce the physical symmetry of the
model in the RVB state, we have made a full symmetry classification of the
Schwinger Boson mean-field ansatz on the square lattice with the projective
symmetry group (PSG) technique.~\cite{psg,psgb,psgc} Then, we have performed
variational Monte Carlo simulations in order to optimize such a Bosonic RVB
state, by using both the permanent Monte Carlo algorithm and the loop gas
algorithm.

We find that the Bosonic RVB wave function gives a rather good
variational description of the system. In addition, we find that the
phase diagram predicted by the DMRG calculations can be well
reproduced. More specifically, the system is found to enter a fully
gapped spin liquid state around $J_{2}=0.4J_{1}$ through a
continuous transition, when the Marshall sign rule in the ground
state begins to be essentially violated. A level crossing of the
spinon excitation is observed around $J_{2}=0.51J_{1}$, when the gap
minimum of the spinon excitation branch is moved from the $\Gamma$
to the $M$ (i.e., $\mathrm{k}=(\pi,0)$) point and a kink appears in
the spin gap as a function of $J_{2}$.

Finally, by PSG symmetry considerations, it can be shown that the spin gap is
always finite at the $M$ point in the spin-liquid region (while it can vanish
at $(\pi,\pi)$, at the transition to the antiferromagnetic phase for small
$J_{2}$). This fact implies that the magnetic structure factor is always
finite at the $M$ point, ruling out a continuous transition to the collinear
phase at large $J_{2}$.

The paper is organized as follows: in Sec.~\ref{sec:model}, we describe the
model and the method; in Sec.~\ref{sec:results}, we present our numerical
results; finally, in Sec.~\ref{sec:conclusion}, we draw our conclusions.

\section{The model and methods}\label{sec:model}

In this paper, we consider the following model:
\begin{equation}
\mathrm{H}=J_{1}\sum_{\langle i,j \rangle}
\vec{\mathrm{S}}_{i}\cdot\vec{\mathrm{S}}_{j}
+J_{2}\sum_{\langle \langle i,j \rangle \rangle}
\vec{\mathrm{S}}_{i}\cdot\vec{\mathrm{S}}_{i},
\end{equation}
where $\langle i,j \rangle$ and $\langle \langle i,j \rangle \rangle$ indicate
nearest-neighbor and next-nearest-neighbor sites on the square lattice,
respectively; $\vec{\mathrm{S}}_{i}$ denotes the spin operator at site $i$.

In the Schwinger Boson representation,~\cite{schwinger} the spin operator is
written as $\vec{\mathrm{S}}=\frac{1}{2}\sum_{\alpha,\beta}
b^{\dagger}_{\alpha}\vec{\sigma}_{\alpha,\beta}b_{\beta}$,
where $b_{\alpha}$ is a Boson operator, $\vec{\sigma}$ is the
Pauli matrix. Bosons should satisfy the no double occupancy constraint
$\sum_{\alpha}b^{\dagger}_{\alpha}b_{\alpha}=1$, in order to be a faithful
representation of the spin-$1/2$ operator. Within this representation, the
Heisenberg super-exchange coupling can be written as (apart from additive
constants)
$\vec{\mathrm{S}}_{i}\cdot\vec{\mathrm{S}}_{j}=-\frac{1}{2}\hat{A}_{i,j}^{\dagger}\hat{A}_{i,j}=\frac{1}{2}\hat{B}_{i,j}^{\dagger}\hat{B}_{i,j}$,
where
$\hat{A}_{i,j}=b_{i\uparrow}b_{j\downarrow}-b_{i\downarrow}b_{j\uparrow}$ and
$\hat{B}_{i,j}=b_{i\uparrow}^{\dagger}b_{j\uparrow}+b_{i\downarrow}^{\dagger}b_{j\downarrow}$.~\cite{schwinger}

In the mean-field treatment, we replace $\hat{A}_{i,j}$ and $\hat{B}_{i,j}$
with their mean-field expectation value $A_{i,j}$ and $B_{i,j}$, so to have:
\begin{eqnarray}
    \mathrm{H}_{\mathrm{MF}}=&-&\frac{1}{2}\sum_{i,j}\left(\Delta_{i,j}\hat{A}_{i,j}^{\dagger}+h.c.\right)\nonumber\\
  &+&\frac{1}{2}\sum_{i,j}\left(F_{i,j}\hat{B}_{i,j}^{\dagger}+h.c.\right)\nonumber\\
  &+&\lambda \sum_{i}
  \left(\sum_{\alpha} b_{i\alpha}^{\dagger}b_{i\alpha}-1\right),
\end{eqnarray}
where $\Delta_{i,j}=J_{i,j}A_{i,j}$, $F_{i,j}=J_{i,j}B_{i,j}$, and
the chemical potential $\lambda$ is introduced to fulfill, on
average, the single-occupancy constraint. The mean-field ground
state has the general form of
\begin{equation}
|\mathrm{MF}\rangle\propto\exp\left\{ \sum_{i,j}a(R_{i},R_{j})(b_{i\uparrow}^{\dagger}b_{j\downarrow}^{\dagger}-b_{i\downarrow}^{\dagger}b_{j\uparrow}^{\dagger})\right\} |0\rangle.
\end{equation}

Then, a suitable RVB wave function in the {\it physical} Hilbert space with
one Boson per site may be obtained by projecting the mean-field state, namely
\begin{equation}\label{eq:RVB}
|\mathrm{RVB}\rangle=\mathrm{P}_{\mathrm{G}}|\mathrm{MF}\rangle,
\end{equation}
where $\mathrm{P}_{\mathrm{G}}$ is a Gutzwiller projector that enforces the
constraint of one Boson per site. The equivalence of the RVB state with the
standard Liang-Doucot-Anderson state~\cite{liang} is clear after projection
onto the physical subspace.

The form of the RVB amplitude $a(R_{i},R_{j})$ is determined by the
parameters $\Delta_{i,j}$, $F_{i,j}$ and $\lambda$. At the mean-field level,
$\Delta_{i,j}$ and $F_{i,j}$ are non-zero only on those bonds with
$J_{i,j} \neq 0$. However, from the variational point of view, we can take
$\{\Delta_{i,j},F_{i,j},\lambda\}$ as a set of free parameters to
construct the RVB state. In such a case, we can also introduce $\Delta_{i,j}$
and $F_{i,j}$ on longer bonds, for which $J_{i,j}=0$.

In order to describe a spin liquid state with the {\it full} symmetry of the
model, the mean-filed parameters $\{\Delta_{i,j},F_{i,j},\lambda \}$ must
satisfy certain symmetry conditions. Since there exists a U(1) gauge degree
of freedom in the Schwinger Boson representation of the spin operator
(i.e., $b_{i,\alpha}\rightarrow b_{i,\alpha}e^{i\phi_{i}} $ leaves
$\vec S_i$ unchanged), the symmetry requirement on the mean-field Hamiltonian
is actually the U(1) gauge projective extension of the physical symmetry of the
model. Such symmetry conditions on the mean-field ansatz can be readily worked
out by the so called PSG technique developed by Wen~\cite{psg} for the
Fermionic representation. The Bosonic version of the PSG is the U(1) subset
of the Fermionic PSG.~\cite{psgb,psgc} Here, we will just point out some basic
structures that are relevant to our study.

\begin{figure}
\includegraphics[width=\columnwidth]{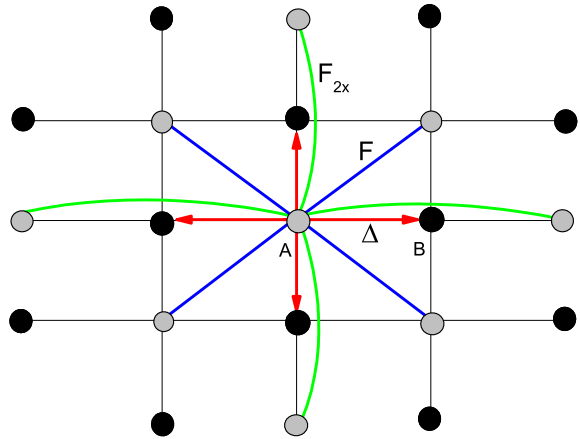}
\caption{\label{fig1}
(Color on-line) An illustration of the mean-field parameters starting
from a site in the sub-lattice A. Gray and dark dots denote sites in
sub-lattices A and B. Here, only bonds up to the third neighbors are reported,
since longer-range parameters are found to be negligibly small after
optimization. The pairing term is always directed from sub-lattice A to
sub-lattice B}
\end{figure}

In the Schwinger Boson formalism, the mean-field parameters $\Delta_{i,j}$
and $F_{i,j}$ describe antiferromagnetic and ferromagnetic local correlations,
respectively (see Appendix~\ref{app1} for the possible phases implied by this
ansatz). Here, we assume a non-zero $\Delta_{i,j}$ between nearest-neighbor
sites. Then, we find that a non-zero $\Delta_{i,j}$ between
next-nearest-neighbor sites is compatible only with the so-called type B
translational property of the mean-field Hamiltonian,~\cite{psg,psgb} which
implies a unit cell with two sites. We find that such state is much higher in
energy than any state in the so-called type A class, characterized by a
manifestly translational invariant mean-field ansatz. Therefore, in the
following we restrict our analysis only to translationally invariant states.
Within the type A states, we find the following general rules for
the mean-field ansatz for a symmetric spin liquid state. First, for
sites belonging to different sub-lattices, only a {\it real} $\Delta_{i,j}$
is allowed. Second, for sites in the same sub-lattice, only a {\it real}
$F_{i,j}$ is allowed. Considering the site $i$ as belonging to A
sub-lattice,~\cite{gauge} the allowed mean-field parameters up to the
fourth-neighbor are given by:
\begin{eqnarray}
F_{i,i+\vec{\delta}_{1}}&=&0, \ \ \ \ \Delta_{i,i+\vec{\delta}_{1}}=\Delta, \\
F_{i,i+\vec{\delta}_{2}}&=&F, \ \ \ \ \Delta_{i,i+\vec{\delta}_{2}}=0, \\
F_{i,i+\vec{\delta}_{3}}&=&F_{2x},\  \Delta_{i,i+\vec{\delta}_{3}}=0, \\
F_{i,i+\vec{\delta}_{4}}&=&0, \
 \ \ \ \Delta_{i,i+\vec{\delta}_{4}}=\Delta_{2xy},
\end{eqnarray}
where $\vec{\delta}_{\mu}$ (with $\mu=1,\dots, 4$) denotes the
vectors connecting the site $i$ to its neighbors, up to the fourth
distance. Here, $\{\lambda,F,\Delta,F_{2x},\Delta_{2xy}\}$ are a set
of real parameters. An illustration of the ansatz used in this study
is shown in Fig.~\ref{fig1}. For the sites $i$ belonging to B
sub-lattice, the sign of $\Delta$ and $\Delta_{2xy}$ should be
reversed (since $\Delta_{i,j}$ is odd by interchanging $i$ and $j$).

\begin{figure}
\includegraphics[width=\columnwidth]{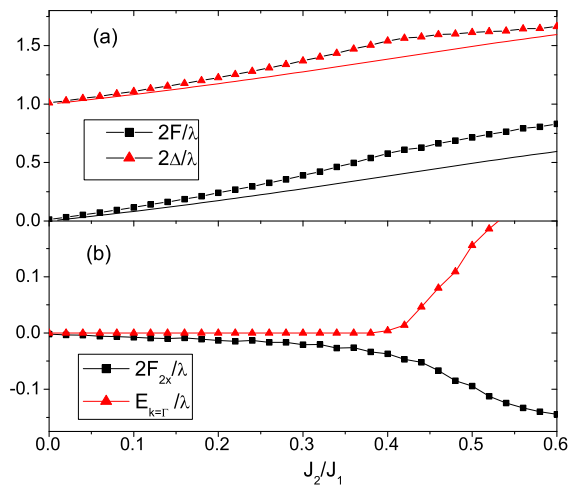}
\caption{\label{fig2}
(Color on-line) The optimized values for various parameters: $2F/\lambda$ and
$2\Delta/\lambda$ (a). The thin lines denote the solution of the mean-field
self-consistent equations. $2F_{2x}/\lambda$ and the (normalized) spinon
gap at the $\Gamma$ point (b).}
\end{figure}

At the mean-field level, both $F_{2x}$ and $\Delta_{2xy}$ are zero, and the
Hamiltonian is given by
\begin{equation}\label{eq:MFhamiltonian}
\mathrm{H}_{\mathrm{MF}}=\sum_{\mathrm{k}\in MBZ
}\psi_{\mathrm{k}}^{\dagger} \left( {\begin{array}{*{20}c}
                                                                      \epsilon_{\mathrm{k}} & 0 & 0 & \Delta_{\mathrm{k}} \\
                                                                      0 & \epsilon_{\mathrm{k}} & -\Delta_{\mathrm{k}} & 0 \\
                                                                      0 & -\Delta_{\mathrm{k}} & \epsilon_{\mathrm{k}} & 0 \\
                                                                      \Delta_{\mathrm{k}} & 0 & 0 & \epsilon_{\mathrm{k}} \\
                                                                    \end{array} } \right)
                                                                    \psi_{\mathrm{k}},
\end{equation}
in which $MBZ$ indicates the reduced (magnetic) Brillouin zone,
$\psi_{\mathrm{k}}^{\dagger}=(b_{A\mathrm{k}\uparrow}^{\dagger},b_{B\mathrm{k}\uparrow}^{\dagger},b_{A\mathrm{-k}\downarrow},b_{B\mathrm{-k}\downarrow})$,
$\epsilon_{\mathrm{k}}=\lambda+2Fg(\mathrm{k})$, and
$\Delta_{\mathrm{k}}=2\Delta\gamma(\mathrm{k})$. Here
$g(\mathrm{k})= \cos(\mathrm{k}_{x})\cos(\mathrm{k}_{y})$,
$\gamma(\mathrm{k})=(\cos(\mathrm{k}_{x})+\cos(\mathrm{k}_{y}))/2$.
The mean-field spectrum is given by
$E_{\mathrm{k}}=\sqrt{\epsilon_{\mathrm{k}}^{2}-\Delta_{\mathrm{k}}^{2}}$
and the minimal spinon gap is given by

\[E_{min}=\left\{
  \begin{array}{ll}
    \sqrt{(\lambda+2F)^{2}-(2\Delta)^{2}}, & \hbox{$2\lambda F<\Delta^{2}$;} \\
    \lambda-2F, & \hbox{$2\lambda
F>\Delta^{2}$.}
  \end{array}
\right.\]
For the first case, the gap minimum is located at the $\Gamma$ point, while
for the second case the gap minimum is at the $M$ point.

Finally, the RVB amplitudes derived from the mean-field ground state are
given by
\begin{equation}\label{eq:amplitude}
a(R_{i}-R_{j})=\frac{1}{N}\sum_{\mathrm{k}\in{MBZ}}\frac{\Delta_{\mathrm{k}}}{\epsilon_{\mathrm{k}}+E_{\mathrm{k}}}e^{i\mathrm{k}\cdot(R_{i}-R_{j})},
\end{equation}
where $N$ is the number of sites, $i\in A$ and $j\in B$. The RVB amplitudes
between sites in the same sub-lattice are identically zero.
We would like to mention that, within the standard formulation based upon
Monte Carlo sampling,~\cite{liang,sandvik,beach} only positive pairing
functions $a(R_{i}-R_{j})$ have been considered so far. In our formulation
this restriction applies only for standard antiferromagnetic phases, while
negative amplitudes are found in the much more interesting spin-liquid phase.

\section{Results}\label{sec:results}

The mean-field Hamiltonian~(\ref{eq:MFhamiltonian}) has been studied by Mila
and collaborators,~\cite{mila} showing that no spin-liquid phases are
stabilized and a direct transition between two ordered phases is present,
with a phase diagram that is very similar to the classical limit.

In order to go beyond this approximation, we now move to the {\it
projected} RVB state of Eq.~(\ref{eq:RVB}), to assess the
possibility that quantum fluctuations may induce a finite spin gap
and, therefore, a stable spin liquid. We thus determine the
parameters in the Bosonic RVB state by optimizing the energy of the
original $J_{1}{-}J_{2}$ model, rather than solving the
self-consistent equations. Then, the spinon gap can be estimated by
inserting back the optimized parameters into the mean-field
dispersion relation $E_{\mathrm{k}}$. Note that the RVB wave
function does not depend on the overall energy scale of the system.
As a result, the spinon gap can be determined only up to a
normalization constant. Here we will use the chemical potential
$\lambda$ as the unit of energy. To have an estimate of the absolute
scale of the spinon gap, we determine the pairing potential $\Delta$
from the equation
\begin{eqnarray}
\Delta=J_{1}\langle\hat{A}_{i,i+x}\rangle=\frac{J_{1}}{N}\sum_{\mathrm{k}\in
MBZ}\frac{\Delta_{\mathrm{k}}\gamma(\mathrm{k})}{E_{\mathrm{k}}}
\end{eqnarray}
by inserting on the right-hand side the optimized values of
$\Delta/\lambda$ and $F/\lambda$, which are $(\Delta/\lambda)_{opt}$
and $(F/\lambda)_{opt}$. Then $\lambda$ can be determined by
requiring that
 $\Delta/\lambda=(\Delta/\lambda)_{opt}$.

The computation of the Bosonic RVB wave function is very expensive
in the Ising basis, since it requires the calculation of permanents,
for which no polynomial algorithm exists.~\cite{permanentbook}
However, on small clusters the calculation is still affordable. In
this work, we have used a $6\times6$ cluster to perform the
optimization of the parameters in the RVB wave function.~\cite{8x8}
It is important to note the key difference between the mean-field
theory and the projected RVB wave function. In the mean-field
theory, the chemical potential $\lambda$ is determined by the
self-consistent equation for the total Boson number. When the spinon
gap approaches zero, the number of Boson will diverge. Thus, on any
finite lattice, the spinon gap can never be zero and a finite-size
gap must exist (see Appendix~\ref{app2} for the details on the
spinon gap in the mean-field approach). On the contrary, after
projection, the constraint of one Boson per site is satisfied {\it
exactly} and such a divergence will not appear. Therefore, the RVB
wave function is well behaved even when the spinon gap is zero. This
fact implies that a vanishing spinon gap can be realized exactly
after optimization of the corresponding projected RVB wave function
on a relatively small cluster.

From our numerical optimization, we find that a spin gap can not be
opened if we keep $\Delta/\lambda$ and $F/\lambda$ only. Moreover,
by a direct optimization of the pairing amplitudes $a(R_{i}-R_{j})$,
a good accuracy can be achieved only by including a third-neighbor
parameter $F_{2x}/\lambda$, while the fourth-neighbor parameter
$\Delta_{2xy}/\lambda$ is found to always negligibly small.
Therefore, in the following, we optimize the wave function with
$\Delta/\lambda$, $F/\lambda$ and $F_{2x}/\lambda$ as variational
parameters. In particular, we find that the inclusion of
$F_{2x}/\lambda$ is crucial for the opening of the spin gap. The
optimized value of the parameters in the RVB wave function are shown
in Fig.~\ref{fig2}.

\begin{figure}
\includegraphics[width=\columnwidth]{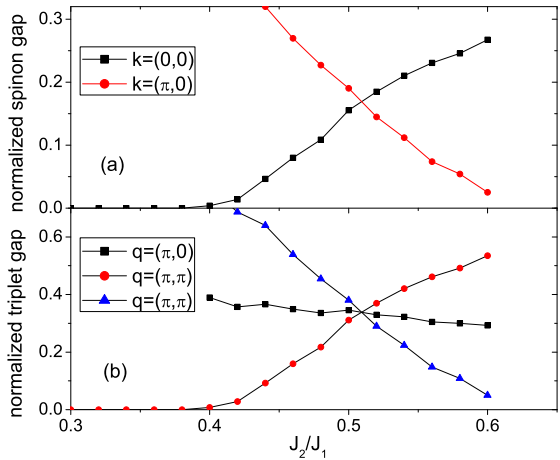}
\caption{\label{fig3}
(Color on-line) The normalized spinon gap at $\Gamma$ and $M$ points in the
spin liquid regime (a). Normalized triplet gap at $\mathrm{q}=(\pi,\pi)$ and
$\mathrm{q}=(\pi,0)$ (b).}
\end{figure}

The spinon gap at the $\Gamma$ and the $M$ points is shown in
Fig.~\ref{fig3}(a). Around $J_{2}=0.51J_{1}$, a level crossing in
the spinon excitation occurs and the gap minimum changes from
$\Gamma$ to $M$. By further increasing $J_{2}$, the spinon gap at
$M$ decreases and eventually approaches zero around
$J_{2}=0.6J_{1}$. At this point the system becomes unstable with
respect to magnetic ordering at $\mathrm{q}=(\pi,0)$. It should be
noted that, although the spinon gap at the $M$ point approaches zero
continuously for $J_{2}=0.6J_{1}$, our state cannot be continuously
connected to the collinear ordered state, and a first-order
transition must exist between the fully symmetric spin liquid and
the collinear ordered magnetic phase. This is clearly seen in the
static spin structure factor:
\begin{equation}
S(\mathrm{q})=\frac{1}{2}\sum_{\mathrm{k}}\left(
\frac{\epsilon_{\mathrm{k}} \epsilon_{\mathrm{q}-\mathrm{k}}
-\Delta_{\mathrm{k}} \Delta_{\mathrm{q}-\mathrm{k}}} {E_{\mathrm{k}}
E_{\mathrm{q}-\mathrm{k}} } -1\right).
\end{equation}
Since $\Delta_{\mathrm{k}=(\pi,0)}=0$ by symmetry (see Appendix~\ref{app3}),
the singularity in the coherence factor
for $E_{\mathrm{k}=(\pi,0)}\rightarrow 0$ is removed and the spin
structure factor at $\mathrm{q}=(\pi,0)$ is always finite. Thus, the
state cannot be connected to the collinear ordered phase, in which
$S(\pi,0)$ diverges. Therefore, we conclude that a first-order
transition must exist between the spin liquid and the collinear ordered phase.

Given the results for the spinon spectrum of Fig.~\ref{fig3}(a), it is possible
to make some prediction on the behavior of the triplet gap as a function of
$J_{2}$. Indeed, to construct a triplet excitation at $\mathrm{q}=(\pi,\pi)$,
we can use two spinons both from the $\Gamma$ point and the $M$
point.~\cite{sublattice} On the contrary, for a triplet excitation with
momentum $\mathrm{q}=(\pi,0)$, we should use one spinon from the $\Gamma$
point and another spinon from the $M$ point. Therefore, the lowest triplet
excitation is always realized at $\mathrm{q}=(\pi,\pi)$ and the energy of
triplet excitation at $\mathrm{q}=(\pi,0)$ is always finite,
see Fig.~\ref{fig3}(b). This is consistent with the result of the static spin
structure factor mentioned above and points to the fact that our spin-liquid
state cannot be continuously connected to the collinear ordered phase.
We note that the peculiar behavior of the triplet excitations found in this
work represents an astonishing consequence of fractionalized spinon excitations
in the spin-liquid phase.

\begin{figure}
\includegraphics[width=\columnwidth]{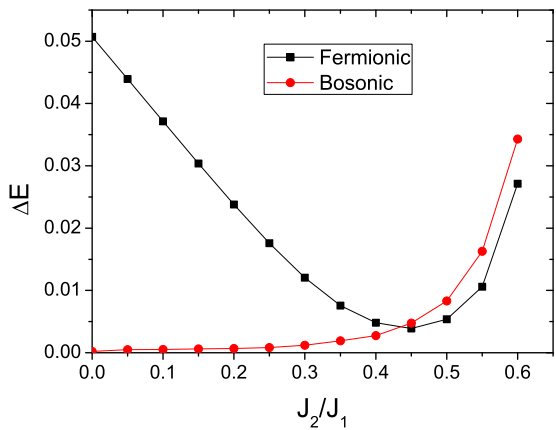}
\caption{\label{fig4}
(Color on-line) Accuracy of the ground-state energy calculated from the best
Fermionic of Ref.~\onlinecite{frvb} and Bosonic RVB variational wave functions
on a $6\times6$ lattice.}
\end{figure}

\begin{figure}
\includegraphics[width=\columnwidth]{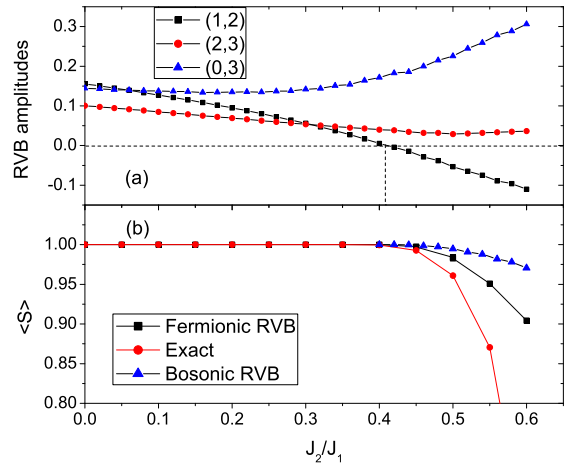}
\caption{\label{fig5}
(Color on-line) The normalized RVB amplitudes $a(R_{i},R_{j})$ on different
distances on a $6\times6$ lattice as functions of $J_{2}/J_{1}$. The
nearest-neighbor $(1,0)$ amplitude has been taken equal to one (a).
The average Marshall sign of Eq.~(\ref{eq:marshall}) calculated from the
Bosonic and Fermionic RVB wave functions and the exact ground state on a
$6\times6$ lattice as functions of $J_{2}/J_{1}$ (b).}
\end{figure}

We would like to mention that our results for the spin gap are quite
similar to the DMRG ones.~\cite{dmrg} Indeed, within both
approaches, the spin gap is found to open around $J_{2}=0.4J_{1}$
and close around $J_{2}=0.62J_{1}$. In addition, a sharp maximum is
present, though its position in the Bosonic RVB approach is found to
correspond to a lower value of $J_{2}$ with respect to the DMRG
study. Moreover, taking the value of $\lambda$ estimated from Eq.11,
which is $\lambda \approx 1.02 J_{1}$ at $J_{2}=0.5J_{1}$, we have
that the maximal spin gap is quite consistent with the DMRG
prediction.

In this work, the sharp maximum in the spin gap is interpreted as the result
of a level crossing in the minimum of the spinon spectrum (from the $\Gamma$
to the $M$ point). In such a picture the lowest triplet excitation within
the symmetric spin liquid phase is always at $\mathrm{q}=(\pi,\pi)$. However,
other possibilities for this structure may exist, among which a spin nematic
liquid phase, which breaks the reflection symmetry $x \to y$ but with all
other physical symmetries intact, is especially interesting.~\cite{parola}
Since the DMRG calculations have been done on rectangular clusters, the
nematic liquid phase can be connected to the symmetric state continuously on
finite lattices.

To further check the accuracy of the Bosonic RVB wave function, we computed
the relative error in the ground-state energy, namely
$\Delta E=|E_0-E_{\rm var}|/|E_0|$, where $E_0$ is the exact ground-state
energy and $E_{\rm var}$ is the variational energy of the RVB state.
In Fig.~\ref{fig4}, we report the accuracy of the Bosonic RVB wave function
on the $6\times6$ cluster, in comparison with the best Fermionic RVB wave
function.~\cite{frvb} For small $J_{2}$, the Bosonic RVB wave function is much
more accurate than the Fermionic RVB wave function, which cannot describe
magnetically ordered states. In this region, our results for the Bosonic wave
function agree with previous calculations reported in
Ref.~\onlinecite{bookbecca}, obtained with a different algorithm~\cite{sandvik}
or a different parametrization.~\cite{beach} For $J_{2} \gtrsim 0.45J_{1}$,
the Fermionic wave function becomes more accurate. However, the error in both
wave functions are similar and both increase with the same trend by increasing
$J_{2}$ up to $J_{2}=0.6J_{1}$.

As pointed out in Ref.~\onlinecite{frvb}, the sign structure of the ground
state is crucial for the origin of the spin liquid phase. For $J_{2}=0$, the
ground-state wave function satisfies the Marshall sign rule.~\cite{marshall}
However, the Marshall sign rule is essentially violated only for
$J_{2} \gtrsim 0.4J_{1}$ and, in the Fermionic RVB approach, a $Z_2$ spin
liquid phase emerges just at the same point.~\cite{frvb} A similar scenario
also appear in the Bosonic representation. In this case, when the RVB
amplitudes from sub-lattice A to sub-lattice B are positive, then the wave
function satisfies the Marshall sign rule, otherwise (if some amplitudes are
negative) the Marshall sign rule is violated. In Fig.~\ref{fig5}, we plot all
the independent RVB amplitudes $a(R_{i},R_{j})$ on a $6\times6$ lattice of the
optimized wave function (with the amplitude between the nearest-neighbor sites
equal to one). For $J_{2}<0.4J_{1}$, all amplitudes are positive and thus the
wave function has the Marshall sign. For $J_{2}>0.4J_{1}$, the amplitude on
bond $(1,2)$ becomes negative and the Marshall sign rule is violated. It is
just at this point that the spin gap opens. Thus, the origin of the spin gap
and the existence of the spin liquid phase can be understood as a result of
violation of the Marshall sign rule. Such an understanding is consistent with
several previous studies,~\cite{topology} in which the topological degeneracy,
which is a hallmark of gapped spin liquid, is argued to be absent in system
satisfying the Marshall sign rule.

Finally, we report in Fig.~\ref{fig5} the average Marshall signs in the
Bosonic and Fermionic RVB wave functions:
\begin{equation}\label{eq:marshall}
\langle S \rangle = \sum_x |\langle x|RVB\rangle|^2
{\rm sign} \left\{ \langle x|RVB\rangle (-1)^{N_\uparrow(x)}\right\},
\end{equation}
where $|RVB\rangle$ denotes the RVB variational state (either Bosonic or
Fermionic) and the sum is over the orthogonal Ising basis $|x\rangle$;
for comparison, we also report the results for the exact ground state, where
$|RVB\rangle$ is replaced by $|\Psi_0\rangle$. The Fermionic RVB wave function
is better in the sense of sign structure and this is consistent with the fact
that the Fermionic wave function has a lower energy for large $J_{2}$.
However, it is clearly seen that both the Bosonic and the Fermionic RVB wave
function underestimate seriously the frustration of the sign in the spin-liquid
regime.

\section{Conclusions}\label{sec:conclusion}

In conclusion, we find the Bosonic RVB wave function generates a
ground-state phase diagram of the $J_{1}-J_{2}$ model on the square
lattice that is qualitatively consistent with DMRG results. A gapped
spin-liquid phase is found for $0.4J_{1}<J_{2}<0.6J_{1}$. The
spin-liquid phase is connected to the staggered magnetic ordered
state through a continuous transition but cannot be connected
continuously to the collinear magnetic ordered state and a
first-order transition between the two must exist. The spin gap is
found to have a maximum around $J_{2}=0.51J_{1}$, as a result of the
level crossing between the spinon at $\Gamma$ and $M$ points. This
fact implies that the lowest triplet excitation is found to be
always at $q=(\pi,\pi)$ in the spin-liquid phase. We also found that
the spin gap opens at the same point where the system violates the
Marshall sign rule. This fact provides strong support for previous
arguments for the absence of topological order in systems satisfying
the Marshall sign rule. Despite that these outcomes are in good
agreement with recent DMRG calculations of Ref.~\onlinecite{dmrg},
the gapless Dirac-type Fermionic RVB ansatz remains slightly more
accurate at the variational level in the highly-frustrated regime.

Tao Li acknowledges financial support by the European Research
Council under Research Grant SUPERBAD (Grant Agreement n.
240524). This work is also supported by NSFC Grant No. 10774187,
National Basic Research Program of China No.2007CB925001 and No.
2010CB923004.

\appendix

\begin{figure}
\includegraphics[width=\columnwidth]{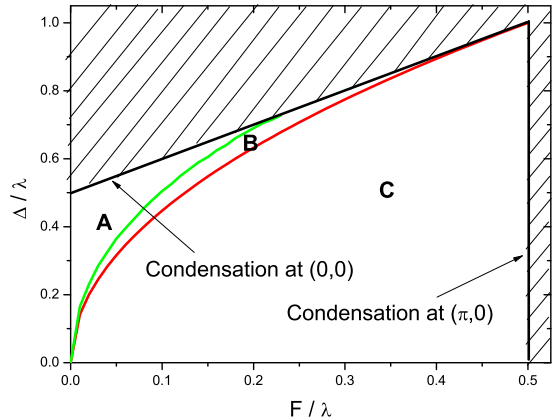}
\caption{\label{fig6}
(Color on-line) Phases described by the Bosonic RVB wave function
$|\mathrm{RVB}\rangle$ in the parameter space $(F/\lambda,\Delta/\lambda )$.}
\end{figure}

\begin{figure}
\includegraphics[width=\columnwidth]{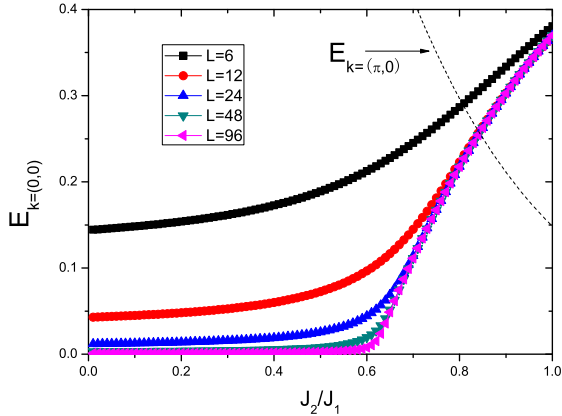}
\caption{\label{fig7}
(Color on-line) The spinon gap predicted by mean-field theory both at $\Gamma$
and $M$ points. For the $\Gamma$ point, we show the scaling of the mean-field
gap with the linear size of the system. The mean-field gap at the $M$ point
in the thermodynamic limit is also reported (dashed line).}
\end{figure}

\section{The various phases described by the wave function studied in
this work}\label{app1}

The various phases described by the wave function studied in this
work are shown in Fig.~\ref{fig6}. Here, we report the various properties as
a function of two parameters, namely $F/\lambda$ and $\Delta/\lambda$.
The case with non-zero $F_{2x}/\lambda$ is qualitatively similar.
The condensation lines denote the magnetically ordered states with staggered
or collinear patterns. The three regions, A, B and C, denote spin-liquid
phases. In regions A and B, the spinon gap minimum is realized at the $\Gamma$
point, while in the region C the gap minimum is moved to the $M$ point.
In region B, the Marshall sign rule is violated while it is satisfied in
region A. The region with slanted lines is physically non accessible.

\section{The mean-field finite size gap}\label{app2}

In the mean-field theory, the spinon is always gapped when the system is
defined on a finite lattice. We report in Fig.~\ref{fig7} the spinon gap
obtained by solving the mean-field self-consistent equations.
Here, we would like to emphasize that the origin of a spinon gap obtained
on finite lattices with the projected Bosonic RVB wave function is totally
different from that obtained within the mean-field approximation. Indeed,
after projection, the number of spinons is fixed (each site is occupied by
one and only one spinon) and the RVB wave function is always well defined.

In fact, we find that the spinon gap is exactly zero for $J_{2}<0.4J_{1}$ from
our optimization on the $6\times6$ lattice, see Fig.~\ref{fig3}. Instead, the
finite size gap in the mean-field theory is much larger and smoother than that
obtained with the projected Bosonic RVB wave function. In addition, we note
that the mean-field theory always predicts a very large gap at the $M$ point
in the thermodynamic limit.

\section{The proof of $\Delta_{\mathrm{k}=(\pi,0)}=0$}\label{app3}

For the ansatz of type A, which is manifestly translational invariant in the
so called uniform gauge, the gauge transformations of the PSG for symmetric
Bosonic spin liquid state is found to be (we have adopted the convention
of Ref.~\onlinecite{psg})

\begin{eqnarray}
G_{P_{x}}&=&\eta_{xP_{x}}^{i_{x}}\eta_{yP_{x}}^{i_{y}}e^{i\phi_{x}}\nonumber\\
G_{P_{y}}&=&\eta_{yP_{x}}^{i_{x}}\eta_{xP_{x}}^{i_{y}}e^{i\phi_{x}}\nonumber\\
G_{P_{xy}}&=&e^{i\phi_{xy}}\nonumber,
\end{eqnarray}
in which $\eta_{xP_{x}},\eta_{yP_{x}}=\pm1$, $\phi_{x},\phi_{xy}=0,\pi/2$.

If we require $\Delta_{i,j}$ to be non-zero between nearest-neighbor sites,
the PSG should satisfy
\begin{eqnarray}
\eta_{xP_{x}}&=&-\eta_{yP_{x}}\nonumber\\
\eta_{xP_{x}}&=&-e^{2i\phi_{x}}\nonumber.
\end{eqnarray}

In the uniform gauge, $\Delta_{i,j}$ is only a function of $R_j-R_i$, so we
can write $\Delta_{i,j}$ as
$\Delta_{(d_{x},d_{y})}$, in which the distance
$(d_{x},d_{y})=(j_{x}-i_{x},j_{y}-i_{y})$. By applying $P_{x}$ and $P_{y}$
successively, we have
\begin{eqnarray}
\Delta_{(-d_{x},-d_{y})}&=&
(\eta_{xP_{x}}\eta_{yP_{x}})^{d_{x}+d_{y}}\Delta_{(d_{x},d_{y})} \nonumber\\
&=& (-1)^{d_{x}+d_{y}}\Delta_{(d_{x},d_{y})} \nonumber.
\end{eqnarray}
However, from the fact that $\Delta_{i,j}=-\Delta_{j,i}$, we have
\[
\Delta_{(-d_{x},-d_{y})}=-\Delta_{(d_{x},d_{y})}.
\]
We thus conclude that $\Delta_{i,j}$ is non-zero only between sites in the
opposite sub-lattices.

To show further that $\Delta_{\mathrm{k}=(\pi,0)}=0$, we need to go to the
sub-lattice uniform gauge.~\cite{gauge} For $\phi_{xy}=0$, the gauge
transformation from the uniform gauge to the sub-lattice uniform gauge is
given by
\[
W_{i}=(-1)^{[\frac{i_{x}+i_{y}}{2}]},
\]
while for $\phi_{xy}=\pi/2$, it is given by
\[
W_{i}=(-1)^{[\frac{i_{x}-i_{y}}{2}]},
\]
in which $[r]$ means the largest integer that is not greater than
$r$. In the sub-lattice uniform gauge, the pairing term
$\Delta_{(d_{x},d_{y})}$ has s-wave symmetry from any site in the A
or B sub-lattice (but has opposite signs for $\Delta_{i,j}$ starting
from the A and B sub-lattices). Thus the total contribution to the
Fourier transform of $\Delta_{(d_{x},d_{y})}$ from distance
$(d_{x},d_{y})$ and all the other symmetry related distances is
proportional to
\[
\Delta_{(d_{x},d_{y})}\left(\cos(\mathrm{k}_{x}d_{x})\cos(\mathrm{k}_{y}d_{y})+\cos(\mathrm{k}_{x}d_{y})\cos(\mathrm{k}_{y}d_{x})\right).
\]
Since $\Delta_{(d_{x},d_{y})}$ is non-zero only when $d_{x}+d_{y}$ is
an odd integer, it is easy to see that
$\Delta_{\mathrm{k}=(\pi,0)}=0$.


\begin{thebibliography}{99}
\bibitem{gauge} Here, we have adopted the sub-lattice uniform gauge rather
than the uniform gauge. The mean-field ansatz in the sub-lattice unifrom gauge
is translational invariant with a doubled unit cell. The gauge transformation
relating uniform gauge and the sub-lattice uniform gauge is given by
$(-1)^{[\frac{i_{x}\pm i_{y}}{2}]}$ depending on the type of PSG, in
which $[r]$ means the largest integer that is not greater than $r$.
\bibitem{8x8} Given the fact that the computational cost for dealing with
permanents grows exponentially with the number of lattice sites, an
optimization procedure on the $8\times8$ cluster is not affordable.
Moreover, an accurate estimation of the gap would require an accuracy on the
energy that is not presently reachable.
\bibitem{sublattice} As a result of the two sub-lattice structure of the
mean-field ansatz, $\mathrm{k}=(0,0)$ is equivalent to $\mathrm{k}=(\pi,\pi)$.
\end{thebibliography}
\end{document}